\documentclass[10pt,aps,prb,twocolumn,showpacs,amssymb,floatfix]{revtex4-1}
\usepackage{amsmath,graphics,epsfig,mathrsfs}
\usepackage{epstopdf}
\usepackage{bm}
\usepackage{color}
\usepackage{soul}
\usepackage{subfigure}
\usepackage{textcomp}

\begin{document}
\title{Spin Transfer Torque in Antiferromagnetic Spin-Valves: From Clean to Disordered Regimes}
\author{Hamed Ben Mohamed Saidaoui}
\affiliation{Physical Science and Engineering Division, King Abdullah University of Science and Technology (KAUST), Thuwal 23955-6900, Kingdom of Saudi Arabia}
\author{Xavier Waintal}
\affiliation{CEA-INAC/UJF Grenoble 1, SPSMS UMR-E 9001, Grenoble F-38054, France}
\author{Aurelien Manchon}\email{aurelien.manchon@kaust.edu.sa}
\affiliation{Physical Science and Engineering Division, King Abdullah University of Science and Technology (KAUST), Thuwal 23955-6900, Kingdom of Saudi Arabia}

\begin{abstract}
Current-driven spin torques in metallic spin-valves composed of antiferromagnets are theoretically studied using the non-equilibrium Green's function method implemented on a tight-binding model. We focus our attention on G-type and L-type antiferromagnets in both clean and disordered regimes. In such structures, spin torques can either rotate the magnetic order parameter coherently (coherent torque) or compete with the internal antiferromagnetic exchange (exchange torque). We show that, depending on the symmetry of the spin-valve, the coherent and exchange torques can either be in the plane, $\propto{\bf n}\times({\bf q}\times{\bf n})$ or out of the plane $\propto{\bf n}\times{\bf q}$, where ${\bf q}$ and ${\bf n}$ are the directions of the order parameter of the polarizer and the free antiferromagnetic layers, respectively. Although disorder conserves the symmetry of the torques, it strongly reduces the torque magnitude, pointing out the need for {\em momentum conservation} to ensure strong spin torque in antiferromagnetic 
spin-valves.
\end{abstract}
\maketitle

\section{Introduction}

The electrical control of the ferromagnetic order parameter through current-driven spin transfer torque \cite{Slonczewski} has enabled the development of promising spin-based devices for memory and logic applications. The core of this phenomenon is the transfer of spin angular momentum from the spin-polarized current to the local magnetic moments. This prediction has been followed by a wealth of experimental discoveries, offering a new breath to spin electronics \cite{review1,review2,chapter}. However, the implementation of spin transfer torque in technologically viable devices requires low current densities of the order of $10^5$ A/cm$^2$ or less, which is difficult to achieve with conventional magnetic materials \cite{hayakawa}. An alternative route towards low current density spin torque devices is the exploitation of {\em antiferromagnets} (AF), rather than ferromagnets (F), as active layers. These devices would take full advantage of the absence of demagnetizing field resulting from the compensation of the magnetic moments. Indeed, the critical current density needed to switch the magnetic state of a ferromagnetic spin-valve reads $j_c\approx\alpha(H_{app}+H_{k}+H_d/2)/\tau$, where H$_d$ is the demagnetizing field, H$_{app}$ is the applied magnetic field and H$_k$ is the anisotropy field, $\tau$ being the spin torque efficiency\cite{Grollier}. Hence, reducing the demagnetizing field (ideally zero in antiferromagnets) would significantly reduce the critical switching current.\par

The first theoretical prediction of an efficient spin torque in metallic spin-valves based on antiferromagnets is due to N\'{u}\~{n}ez et al\cite{Nunez}. These spin-valves comprise two antiferromagnets separated by a normal metal (N), of the form AF/N/AF as depicted in Fig. \ref{fig:fig1}. Using a ballistic tight-binding model in the absence of defects or impurities, the authors proposed that the {\em staggered} spin density produced in the antiferromagnetic reference layer is transmitted to the antiferromagnetic free layer and exerts a torque on the order parameter, of the form ${\bf T}=T_{\parallel}{\bf n}\times({\bf q}\times {\bf n})$ where ${\bf n}$ and ${\bf q}$ are the order parameters of the free and reference antiferromagnets, respectively. Together with the spin torque, antiferromagnetic spin-valves are also expected to display magnetoresistance. These results have been confirmed by {\em ab initio} calculations on  Cr(100)/Au(100)/Cr(100) \cite{haney2007} and $\gamma$-FeMn/Cu/$\gamma$-FeMn \cite{xu} stacks. 
The current-driven 
order parameter dynamics has been investigated within the macrospin approximation accounting for two coupled sub-lattices and uncovering both angular and linear oscillations of the order parameter\cite{Gomonay}. Finally, the control of antiferromagnetic {\em domain walls} by external currents \cite{Swaving2011,Swaving2012,Hals,afdwsw,ren} has recently attracted increasing attention. The spin torque has a form similar to the one derived in ferromagnetic domain walls, ${\bf T}=b_J({\bf u}\cdot{\bf \nabla}){\bf n}-\beta b_J{\bf n}\times({\bf u}\cdot{\bf \nabla}){\bf n}$ ($b_J$ being the torque magnitude, $\beta$ the non-adiabaticity parameter and ${\bf u}$ the current direction). Although the actual time-dependent current-driven dynamics of the antiferromagnetic domain walls remains to be explored in details, a number of important aspects have been theoretically uncovered such as the initial and terminal drift velocity of the walls, and the fact that it acquires a non-equilibrium magnetization. It has also been recently predicted that spin waves could be used to move these antiferromagnetic domain walls \cite{afdwsw}.\par

An important hurdle to detect current-induced spin torques in antiferromagnetic spin-valves remains the vanishingly small magnetoresistance of such systems \cite{wang}. To circumvent this issue, Wei et al. \cite{Tsoi} and Urazhdin and Anthony \cite{Urazhdin} have systematically studied the effect of a spin-polarized current on the exchange bias between the antiferromagnetic layer and the polarizing ferromagnet in a regular exchange-biased ferromagnetic spin-valve of the form AF/F/N/F. Although this structure is notably different from the antiferromagnetic spin-valve studied by Nunez et al.  \cite{Nunez}, evidence of current-driven exchange bias modifications has been observed. Nevertheless, the complexity of this structure renders the analysis of the magnetization hysteresis quite challenging. The recent groundbreaking observation of giant tunneling anisotropic antiferromagnetic magnetoresistance in IrMn-based tunnel junctions by Park et al. \cite{Wunderlich} opened new avenues in this area by allowing for the disentanglement between the spin injection and the detection of the magnetization dynamics in antiferromagnetic spin-valves. The further realization of a memory device based on epitaxially-grown antiferromagnets by Marti et al.\cite{marti1,marti2} paves the way to antiferromagnetic spintronics.\par

Whereas the existence of spin torque in antiferromagnetic spin-valves has been confirmed by several numerical calculations, most of these studies address ballistic transport in the {\em clean} limit (i.e. in the absence of momentum scattering by defects or impurities) of a 1-dimensional antiferromagnetic chain. However, in an antiferromagnet the {\em staggered} magnetic texture responsible for the spin polarization of the itinerant electrons varies on the scale of the crystal unit cell. Therefore one expects {\em momentum conservation} to be a crucial ingredient to obtain sizable torques in such structures.\par

In the present work, we investigate the impact of disorder-induced momentum scattering on the spin torque present in metallic antiferromagnetic spin-valves. We theoretically investigate the nature of spin torque from clean to disordered regimes in different combinations of G-type and L-type antiferromagnets (depicted in Fig. \ref{fig:fig1}). In G-type antiferromagnets, each magnetic site is aligned antiferromagnetically with its surrounding nearest-neighbors, while in L-type antiferromagnets, the system is composed of ferromagnetically magnetized monolayers which are antiferromagnetically aligned with each other. We demonstrate that (i) the torque can be classified into two types, {\em coherent} and {\em exchange} torques. The former tends to rotate the neighboring spins together clockwise (or anticlockwise) promoting a coherent rotation of the order parameter ${\bf n}$ while the second one tends to make them pointing in opposite directions and hence competes with the antiferromagnetic exchange; (ii) the coherent in-plane torque dominates in symmetric spin-valves (composed of the same type of antiferromagnets), as predicted by N\'{u}\~{n}ez et al. \cite{Nunez}, whereas the coherent perpendicular torque dominates in asymmetric spin-valves (composed 
of two different antiferromagnets); (iii) momentum scattering and disorder dramatically damages the spin torque magnitude demonstrating the importance of momentum conservation in order to achieve efficient spin torque in antiferromagnetic spin-valves.\par

Our work is divided as follows. Section \ref{sec:2} presents the theoretical method and offers a general discussion about the nature of the spin transport and the torque in antiferromagnets. The numerical results for the clean and disordered regimes are presented and discussed in Section \ref{sec:3}. Conclusion and perspectives are provided in Section \ref{sec:4}.

\section{Background \label{sec:2}}
\subsection{Theoretical method}
\begin{figure}[h!]
  \centering
  \includegraphics[scale=0.55]{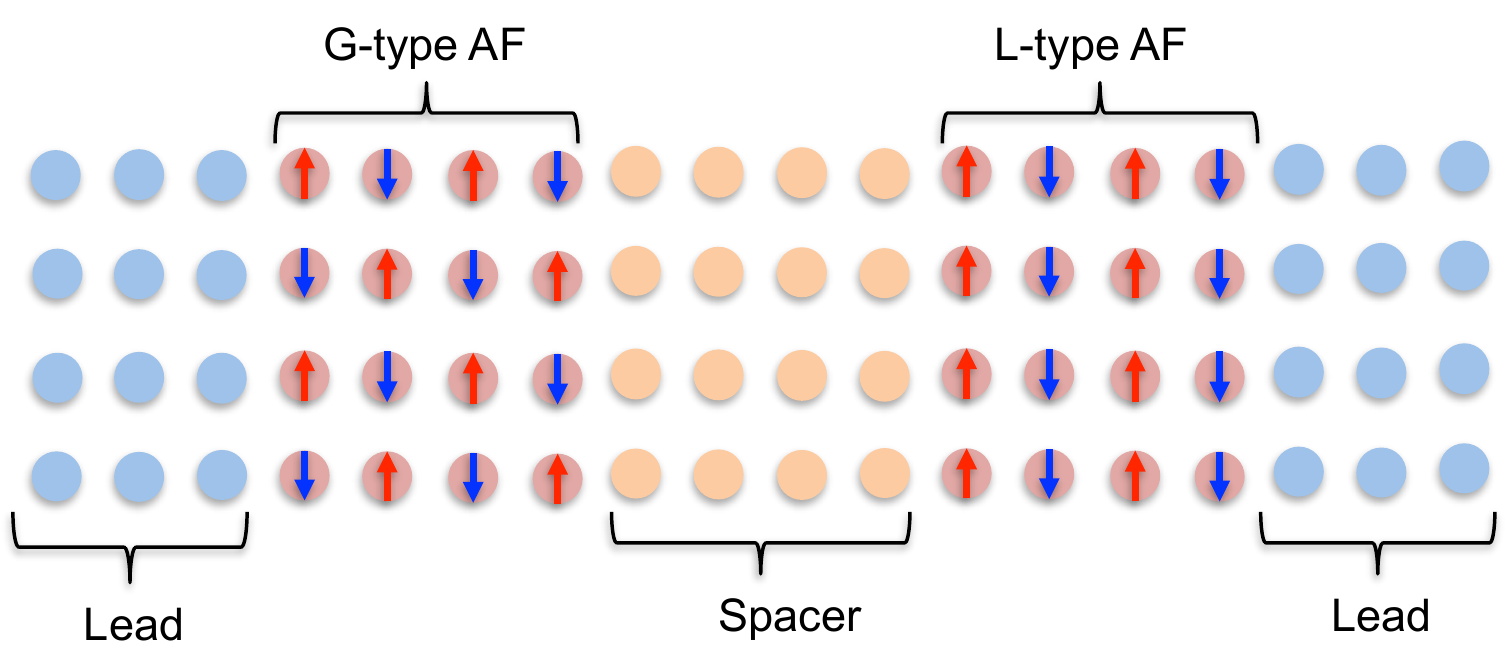}
  \caption{(Color online) Square lattice representation of an antiferromagnetic spin-valve, as implemented in the tight-binding code. The systems studied in this work are composed of a combination of G-type and L-type antiferromagnets separated by a metallic spacer and connected to two external leads.\label{fig:fig1}}
\end{figure}
The multilayer stack is modeled by a two-dimensional square lattice composed of magnetic and non-magnetic layers connected to left and right electrodes by the intermediate of non-magnetic leads, as illustrated in Fig. \ref{fig:fig1}. The spin transport is investigated using Keldysh non-equilibrium Green's functions technique\cite{Rocha,Garcia} (NEGF), as implemented in the KNIT code \cite{Xavier}. We use a single band tight binding approximation which extends all physical quantities in the basis of the local atomic sites wave function. In this model, the Hamiltonian reads

\begin{eqnarray}
&&{\hat H}=\sum_{i}\epsilon_{i}{\hat c}_{i}^+ {\hat c}_{i}-\sum_{i,i'}t_{i,i'}{\hat c}_{i}^+ {\hat c}_{i'} - \frac{\Delta}{2} \sum_{i}{\hat c}_{i}^+{\bm\Omega}_{i}\cdot{\hat{\bm\sigma}}{\hat c}_{i}
\end{eqnarray}
Where $\epsilon_i$ is the on-site energy, $t_{i,i'}$ is the hopping parameter between the sites $i$ and $i'$, restricted to nearest neighbors and $\Delta$ is the exchange energy between the local moment ${\bm \Omega}_i$ on site $i$ of the scattering region and the itinerant electron spin. $\hat{\bm \sigma}$ is the vector of spin Pauli matrices where $\hat{}$ denotes a 2$\times$2 matrix in spin space, ${\hat c}_{i}^+$ is the creation operator of an electron on site $i$ such that ${\hat c}_{i}^+=(c_{i\uparrow}^+,c_{i\downarrow}^+)$, where $\uparrow,\downarrow$ refers to the spin projection along the quantization axis. Notice that the local magnetic moment direction ${\bm \Omega}_i$ on site $i$ is determined by the magnetic nature of the layer: it is uniform for ferromagnets and staggered for antiferromagnets.\par

The non-equilibrium transport properties of the metallic spin-valve are determined through the conventional NEGF method. The conductance for electrons at energy $\epsilon$ is given by Landauer formula, $g=\frac{e^2}{h} {\hat G}^R(\epsilon)\Gamma_{\rm L}(\epsilon)({\hat G}^R(\epsilon))^+\Gamma_{\rm R}(\epsilon)$ where ${\hat G}^R(\epsilon)$ is the retarded Green's function and ${\hat \Gamma}_{\rm L,R}(\epsilon)=\Im\left[\Sigma_{\rm L,R}(\epsilon)\right]$ describes the electron lifetime in the left (right) lead and $\Sigma_{\rm L,R}(\epsilon)$ is the self energy term which introduces the interaction of the central system with the leads. The local electron density matrix at energy $\epsilon$ on site $i$ is given by the imaginary part of the lesser Green's function ${\hat{\bm\rho}}_i=\frac{1}{2\pi}\Im {\int{{\hat G}^<_{ii}(\epsilon)d\epsilon}}$. The latter is related to the retarded Green's function as ${\hat G}^<(\epsilon)={\hat G}^R(\epsilon) {\hat \Sigma}^< (\epsilon)({\hat G}^R(\epsilon))^+$, where ${\hat \Sigma}^<(\epsilon)$ is the lesser self energy that describes the effect of the two leads on the scattering region: ${\hat \Sigma}^<(
\epsilon)=i{\hat \Gamma}_{\rm L}(\epsilon)f_{\mu_{\rm L}}+i{\hat \Gamma}_{\rm R}(\epsilon)f_{\mu_{\rm R}}$. A bias voltage $\mu_{\rm L}-\mu_{\rm R}=eV$ is applied across the system, ensuring the continuous flow of current from the left to the right lead. In this work, we focus on non-equilibrium properties induced by the bias voltage $eV$. Therefore, the non-equilibrium spin density is defined as $\delta{\bm s}_i=\text{Tr}[\hat{\bm \sigma}\delta\hat{\bm \rho}_i]$ where $\delta\hat{\bm \rho}_i=\partial_{eV}\hat{\bm \rho}_ieV$ is the non-equilibrium density matrix induced by the voltage $eV$ and $\text{Tr}$ denotes the trace operator. Similarly, the local spin torque is defined as the torque exerted by the non-equilibrium spin density  $\delta{\bm s}_i$ on the local moment of site $i$
\begin{equation}\label{eq:torque}
{\bm \tau}_i=\Delta{\bm \Omega}_i\times\delta{\bm s}_i.
\end{equation}
The torque in Eq. (\ref{eq:torque}) is defined locally and extends over the layers area. In the remaining of this article and in order to address the electrical efficiency of the spin torque, the total torque exerted on the magnetic layer is normalized to the flowing current,
\begin{equation}
{\bf T}=\frac{1}{geV}\sum_i{\bf \tau}_i
\end{equation}
where the summation runs over all the sites of the (ferro- or antiferro-)magnetic layer subject to the torque. Similarly, the non-equilibrium spin density is normalized to the non-equilibrium electron density, in order to obtain comparable efficiencies for different magnetic structures. Therefore, the spin density is unitless and the spin torque is given by a torque efficiency in units of $h/e$.

\subsection{Staggered spin polarization from antiferromagnets}
 
The spin transfer torque occurring in ferromagnetic spin-valves is a result of the transfer of spin angular momentum from the incoming spin current, polarized along the magnetization ${\bf p}$ of the polarizing layer, to the local magnetic moments of the free layer aligned along ${\bf m}$. The resulting torque in metallic ferromagnetic spin-valves is of the form\cite{Slonczewski} ${\bf T}^0_{\rm F}=a_{\rm J}{\bf m}\times({\bf p}\times{\bf m})$. An extension of this torque to ferromagnetic domain walls yields\cite{review1,review2,chapter} ${\bf T}_{\rm F}^{\rm DW}=b_{\rm J}(1-\beta{\bf m}\times)({\bf u}\cdot{\bm\nabla}){\bf m}$. As mentioned in the introduction, numerical and phenomenological evaluations of the spin torque in {\em symmetric} antiferromagnetic spin-valves gives the general form\cite{Nunez} ${\bf T}_{\rm AF}^0=a_{\rm J}'{\bf n}\times({\bf q}\times{\bf n})$ while in antiferromagnetic domain walls\cite{Swaving2011,Swaving2012,Hals} ${\bf T}_{\rm AF}^{\rm DW}=b'_{\rm J}(1-\beta{\bf n}\times)({\bf 
u}\cdot{\bm\nabla}){\bf n}$, where ${\bf q}$ and ${\bf n}$ are the order parameters of the polarizing and free antiferromagnets, respectively. The similarity between the torques ${\bf T}^{0,\rm DW}_{\rm F}$ and  ${\bf T}^{0,\rm DW}_{\rm AF}$ suggests that an {\em antiferromagnetic spin polarization} defined by the order parameter ${\bf q}$ emerges in antiferromagnetic spin-valves and plays a role equivalent to the ferromagnetic spin polarization in ferromagnetic spin-valves. This antiferromagnetic spin polarization has been identified by Nunez et al.\cite{Nunez} as a non-equilibrium {\em staggered} polarization defined by the magnetic order of the antiferromagnet. \par

\begin{figure}[h!]
  \centering
  \includegraphics[scale=0.4]{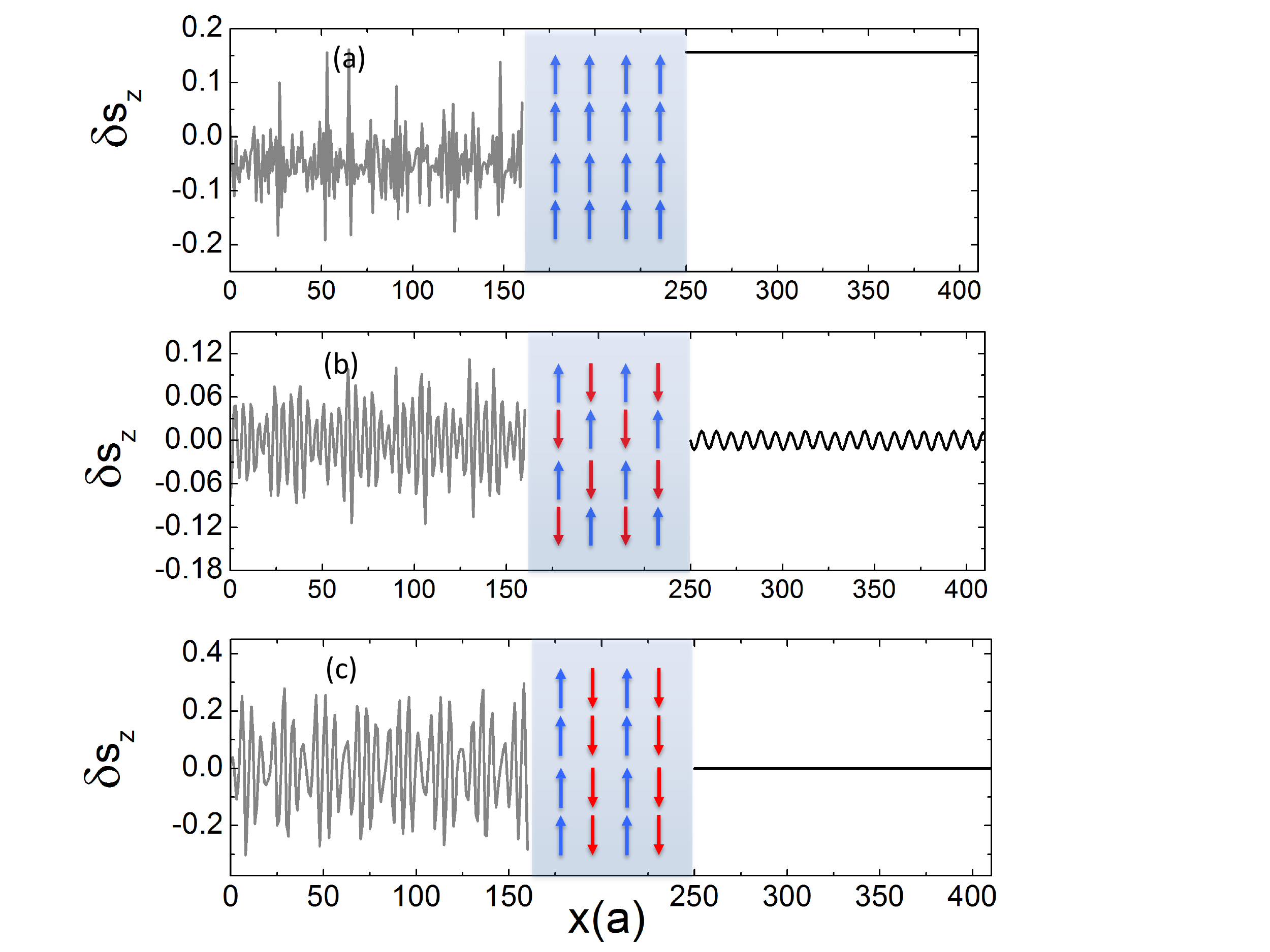}    	
  \caption{Spin density profile generated by itinerant electrons flowing in (a) a ferromagnet, (b) a G-type antiferromagnet and (c) L-type antiferromagnet. The left (right) hand side plots show the reflected (transmitted) spin densities. The calculations were done with the parameters: $\epsilon=-3$ eV and $\Delta/t=1$ \label{fig:fig2}}
\end{figure}

 In order to visualize explicitly such a {\em staggered} spin polarization, we calculated the spatial distribution of the spin density when a current flows in a single ferromagnet [Fig. \ref{fig:fig2}(a)], a G-type antiferromagnet [Fig. \ref{fig:fig2}(b)]  and an L-type antiferromagnet [Fig. \ref{fig:fig2}(c)]. For this calculation, the magnetic layers 
and the non-magnetic leads are composed of an array of 20$\times$20 and 200$\times$20 sites, respectively. The left (right) panels in Fig. \ref{fig:fig2} represent the spatial profile of the spin density in the left (right) lead, which corresponds to the reflected (transmitted) spin density.\par

As expected, the ferromagnetic layer filters the incoming electrons and yields a spin polarization that is spatially invariant [see Fig. \ref{fig:fig2}(a)]. Note that the reflected spin density displays a rather complex spatial profile, due to the interferences between leftward and rightward electron waves. In comparison, the spin filtering through the G-type antiferromagnet displays a regular staggered spin texture, slightly modulated by an envelope due to the quantum confinement along the transport direction within the right lead [Fig. \ref{fig:fig2}(b)]. In contrast with the ferromagnetic case, the reflected spin density displays a much more coherent staggered texture combining spin-dependent wave interference and quantum confinement in the left lead. The transmitted staggered spin density constitutes a {\em pseudo spin} polarization that survives away from the antiferromagnet and is responsible for the torque in G-type antiferromagnetic spin-valves, as proposed by Nunez et al.\cite{Nunez}. Interestingly, 
the spin density profile emerging from the L-type antiferromagnet displayed in Fig. \ref{fig:fig2}(c) is quite intriguing. Whereas it produces a coherent staggered spin texture by reflection, similar to the G-type antiferromagnet, no net (pseudo) spin polarization arises from transmission. \par

These calculations uncover an important aspect of spin transport in antiferromagnets: the effective {\em antiferromagnetic spin polarization} is strongly dependent on the antiferromagnetic order and different characteristics are expected in different antiferromagnetic materials, such as G-, A-,C- and L-type and non-collinear structures. In the special case of collinear antiferromagnets considered in the present work, whereas G-type antiferromagnets can create a staggered spin polarization in both transmission and reflection, L-type antiferromagnet can only produce a staggered spin polarization in reflection. Furthermore, the staggered polarization obtained through reflection in L-type antiferromagnets is larger than in G-type antiferromagnets. This emphasizes the importance of the multiple spin-dependent reflections that take place in the metallic spacer of the antiferromagnetic spin-valves, as well as the importance of the spatial coherence of the magnetic texture (see also Ref. 7).\par

\section{Numerical results \label{sec:3}}
\subsection{Definition of the torque}

\begin{figure}
\centering
\includegraphics[width=0.8\columnwidth]{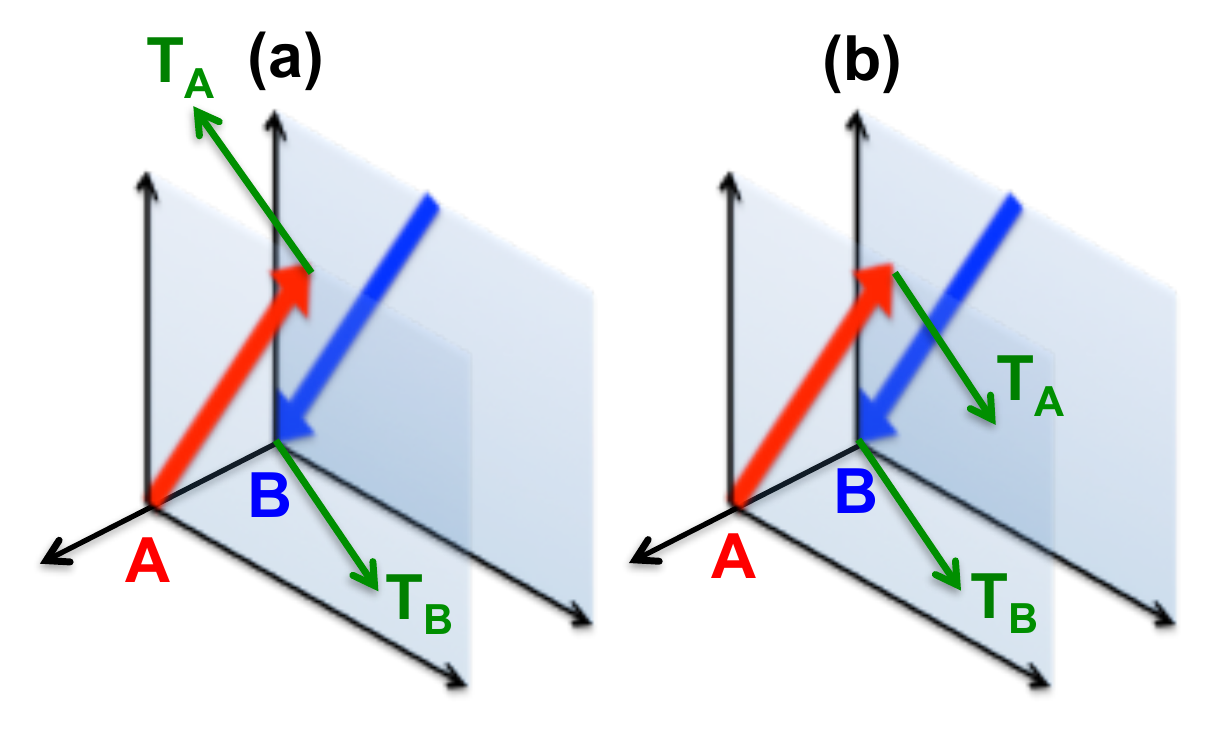}    	
\caption{(Color online) Schematics of the {\em coherent} (a) and {\em exchange} torques (b) (green arrows) applied on the spin of two neighboring sites A and B (red and blue arrows, respectively). In the case of coherent torque, the torques exerted on the two neighbors are opposite to each other, which results in a coherent rotation of the order parameter. In the case of  exchange torque, the torques exerted on the two neighbors are in the same direction, which results in breaking the collinear ordering of the spins.\label{fig:fig3}}
\end{figure}

The present work is restricted to collinear antiferromagnets composed of two sublattices, denoted A and B which feel a different torque, ${\bf T}_{\rm A}$ and ${\bf T}_{\rm B}$, as illustrated in Fig. \ref{fig:fig3}. If the torques exerted on the two sublattices are opposite in sign [Fig. \ref{fig:fig3}(a)], the two sublattices rotate {\em coherently} resulting in the reorientation of the order parameter of the antiferromagnet. When the two torques have the same sign [Fig. \ref{fig:fig3}(b)], the two sublattices tend to misalign with each other. The former, referred below as the {\em coherent} torque, is useful for the electrical control of the order parameter of the antiferromagnet and triggers coherent current-induced dynamics, as studied by Gomonay and Loktev \cite{Gomonay}. The latter, referred to as {\em exchange} torque, competes with the antiferromagnetic exchange and is therefore expected to have a negligible impact on the magnetization dynamics due to the usually very large magnitude of the 
antiferromagnetic exchange. In general, since the local torque varies spatially in 
magnitude and direction, both {\em coherent} and {\em exchange} torques are present. Therefore, for the sake of completeness, both types of torques will be addressed in the numerical simulations. From symmetry considerations, the two types of torques adopt the general form ${\bf T}=T_\|{\bf n}\times({\bf q}\times{\bf n})+T_\bot {\bf n}\times{\bf q}$, where ${\bf n}$ and ${\bf q}$ are the order parameters of the polarizer and free (ferro- or antiferro-)magnetic layers, respectively. The first term is referred to as {\em in-plane} torque and the second term is called {\em out-of-plane} torque.

\subsection{Spin torques in the clean limit}
\begin{figure}
\centering
\includegraphics[width=1.2\columnwidth]{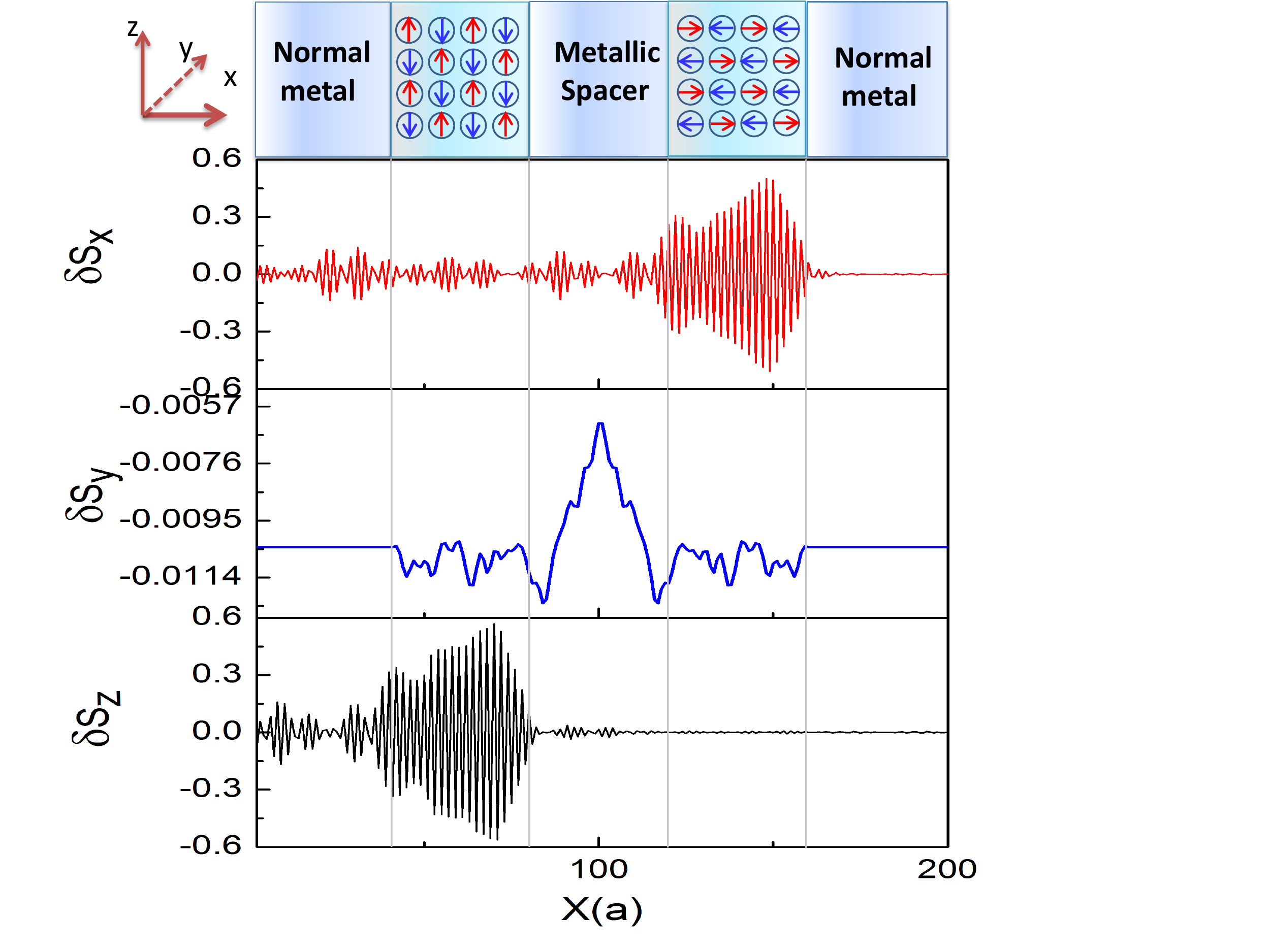}	
\caption{(Color online) Spatial profile of the three components of the non-equilibrium spin density obtained for a symmetric G-type spin valve. The left AF layer is oriented along {\bf z} whereas its right AF layer is oriented along {\bf x}. The parameters are the same as in Fig. \ref{fig:fig2}.\label{fig:fig4}}
\end{figure}
In the present section,we consider a spin-valve composed of two antiferromagnets (either G-type or L-type) separated by a metallic spacer. Each layer contains 400 atoms distributed on a $20\times20$ square lattice of alternating spins. As mentioned in the previous section, the torque is divided by the conductance and the applied bias and is expressed in units of $h$/e. In our calculation we used an electron energy $\epsilon = $-3 eV, the exchange interaction and the hopping energy are chosen such as $\Delta/t = 1$.\par

In order to illustrate the rich transport occurring in antiferromagnetic spin-valves, let us first consider a symmetric G-type spin-valve, whose left AF layer is oriented along {\bf z} whereas its right AF layer is oriented along {\bf x}. Fig. \ref{fig:fig4} represents the spatial distribution of the three components of the non-equilibrium spin density when the electric charges flow from left to right. Consistently with the discussion proposed in the previous section, the spin density gets polarized in the first AF layer along {\bf z} and acquires a staggered texture along this direction. The magnitude of the {\bf z}-component of the spin density injected into the right AF layer is actually quite small, but survives over the volume of the layer. Reciprocally, the staggered spin density polarized along {\bf x} originating from the right layer is reflected back to the left AF layer. Its magnitude is much larger, due to the strong spin polarization obtained through reflection as discussed previously [see Fig. \ref{fig:fig2}(b)]. \par

\begin{figure}
\begin{center}
  \includegraphics[width=1.\columnwidth]{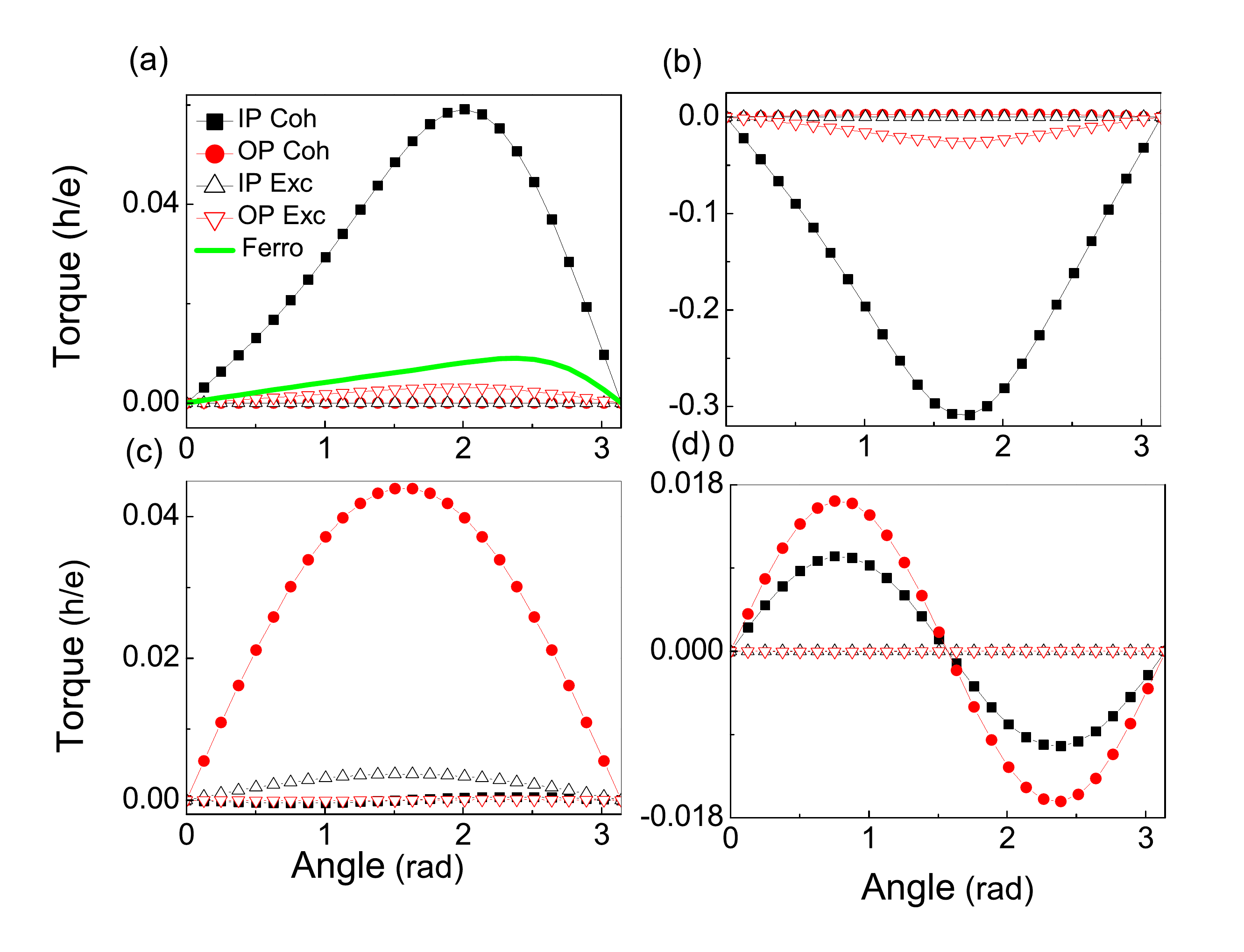}    	
  \caption{(Color online) Dependence of the efficiency of the spin torque exerted on the right layer when electrons are flowing from left to right as function of the raltive angle between the two layer's magnetic moments. The structures calculated are symmetric antiferromagnetic spin-valves composed of (a) G-type and (b) L-type antiferromagnets, as well as asymmetric spin-valves composed of L-type/G-type (c) and G-type/L-type (d). In-plane ($\Box$) and perpendicular components ($\bigcirc$) of the coherent torque as well as in-plane ($\triangle$) and perpendicular components ($\bigtriangledown$) of the exchange torques are represented. The solid line in (a)  presents the in-plane coherent torque calculated in conventional symmetric ferromagnetic spin-valves showed for reference. The parameters are the same as in Fig. \ref{fig:fig2}.\label{fig:fig5}}
  \end{center}
\end{figure}

This complex spin texture has direct implications in terms of spin torque, as displayed in Fig. \ref{fig:fig5}(a). The components of the spin density along {\bf z} and {\bf x} [Fig. \ref{fig:fig2}(a) and (b)] result in a strong in-plane coherent torque and weaker out-of-plane exchange torque, respectively. The precession of the itinerant spins around the local moments of the right antiferromagnet generates a component of the non-equilibrium spin density along  {\bf y}. This additional {\bf y}-component is expected to produce an additional out-of-plane coherent torque as well as an in-plane exchange torque. However, the magnitude of the {\bf y}-component of the non-equilibrium spin density being vanishingly small results in negligible torques [Fig. \ref{fig:fig5}(a)]. As a reference, the torque obtained for the conventional ferromagnetic spin-valve is reported in Fig. \ref{fig:fig5}(a) (solid line). The tight-binding parameters and dimensions of the system are the same as in the symmetric G-type 
antiferromagnetic spin-valve studied above, we only imposed the local magnetization of the polarizing and free magnetic layers to align ferromagnetically. As initially observed by Nu\~nez et al. \cite{Nunez}, for the same set of parameters, the spin torque in the antiferromagnetic spin-valve is about one order of magnitude larger than the spin torque in the ferromagnetic spin-valve. This difference is attributed to the fact that spin torque in antiferromagnets spreads over the {\em bulk} of the magnetic layer, whereas the spin torque in ferromagnets is confined to the interface. \par

We now extend our investigation to a comparative estimation of the torque magnitude in three types of antiferromagnetic spin-valves: symmetric (G-type/G-type and L-type/L-type) [Fig. \ref{fig:fig5}(a) and (b)] and asymmetric (G-type/L-type) [Fig. \ref{fig:fig5}(c) and (d)]. Notice that while all the magnetic systems studied here have the same set of tight-binding parameters, they differ from each other by their local magnetic configuration. First we note that symmetric (G-type, L-type and ferromagnetic) spin-valves present a dominating in-plane coherent torque and perpendicular exchange torque. Second, we observe that the torque calculated in the L-type antiferromagnetic spin-valve is one order of magnitude larger than the one calculated in the G-type spin-valve. It is also worth noticing that the angular dependence is not a simple $\sin(\theta)$ but a deviation is visible, related to the multiple spin-dependent reflections in the metallic spacer, as in the well-known case of ferromagnetic spin-valves.\par

\begin{figure}
\centering
\includegraphics[width=0.8\columnwidth]{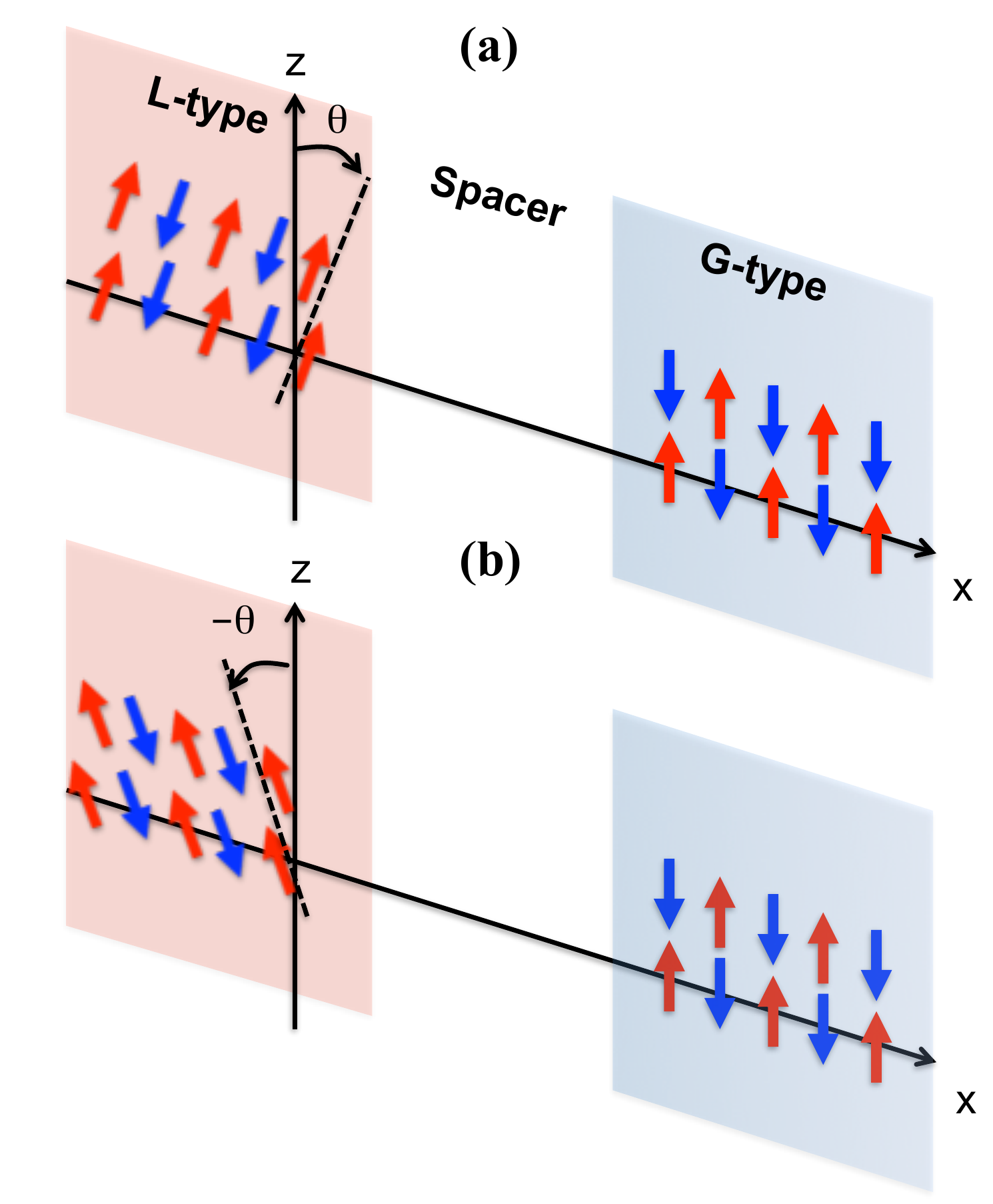}  	
\caption{(Color online) Illustration of an asymmetric antiferromagnetic spin-valve composed of L-type and G-type layers stacked along the x-axis. (a) The order parameter of the G-type antiferromagnet is aligned along the z-axis, while the order parameter of the L-type antiferromagnet is rotated by an angle $\theta$ around the x-axis. (b) The rotation of the magnetic moments of both antiferromagnets by an angle $\pi$ around the z-axis is equivalent to rotating the order parameter of the L-type antiferromagnet by an angle Ð$\theta$ around the x-axis.\label{fig:lg}}
\end{figure}

The spin torque calculated for asymmetric antiferromagnetic spin-valves displays completely different characteristics compared to the symmetric spin-valves. When the spin torque is exerted from the L-type antiferromagnet on the G-type antiferromagnet [Fig. \ref{fig:fig5}(c)], the coherent torque is dominated by the perpendicular component, whereas the in-plane component vanishes. Similarly, the exchange torque is dominated by its in-plane component. This situation is simply the opposite of what has been found in symmetric spin-valves, as discussed above. Most intriguingly, when the torque is exerted by the G-type antiferromagnet on the L-type antiferromagnet [Fig. \ref{fig:fig5}(d)], the coherent torque possesses both an in-plane and a perpendicular components, whereas no exchange torque is observed. Furthermore, the angular dependence becomes radically different, displaying a $\sin(2\theta)$-dependence. This angular dependence can be understood using a phenomenological reasoning developed by Haney and MacDonald for F/N/AF spin-valves\cite{Haney}. Let us consider the L-type/G-type spin-valve illustrated in Fig. \ref{fig:lg}. The order parameter of the G-type antiferromagnet is aligned along {\bf z}, while the order parameter of the L-type antiferromagnet is rotated by an angle $\theta$ around {\bf x} [Fig. \ref{fig:lg}(a)]. The rotation of the magnetic moments of both antiferromagnets by an angle $\pi$ around {\bf z} is equivalent to rotating the order parameter of the L-type antiferromagnet by an angle -$\theta$ around {\bf x} [Fig. \ref{fig:lg}(b)]. Therefore, the torque must be an odd function of the angle $\theta$ and can be Fourier-expanded as a summation of $\sin(n \theta)$. In addition, since the G-type antiferromagnet itself is invariant under a rotation of $\pi$ along {\bf x}, $\sin(n(\theta +\pi))=\sin(n\theta)$, which restricts the value of $n$ to $n=2p$ resulting in the $\sin(2\theta)$-dependence of the torque exerted by the G-type antiferromagnet on the L-type antiferromagnet. On the other hand, unlike the G-type antiferromagnet (which is invariant under a $\pi$-rotation), the L-type antiferromagnet is only invariant under a rotation of 2$\pi$ around {\bf x} and then only a $\sin(\theta)$-dependence of the torque exerted by the L-type antiferromagnet on the G-type antiferromagnet is obtained. It is worth noticing that the $\sin(2\theta)$-dependence of the torque obtained in the L-type/G-type spin-valve is not related to the wavy texture of the spin torque reported by Boulle et al. \cite{boulle}, which is controlled by the spatial dependence of the spin accumulation. This phenomenon is absent from our calculations since no spin relaxation is implemented.

We also 
find that the efficiency of the torque exerted in both symmetric and asymmetric antiferromagnetic spin-valves is at least one order of magnitude larger than the one exerted in the ferromagnetic spin-valves for the same set of parameters. These calculations indicate that the nature of spin torques in antiferromagnetic spin-valves dramatically depends on the magnetic configuration of the antiferromagnetic layers. This renders the experimental search for spin torque even more challenging, particularly when noticing that the present theoretical work is limited to {\em collinear} antiferromagnets. \par
\begin{figure}
\begin{center}
  \includegraphics[width=0.9\columnwidth]{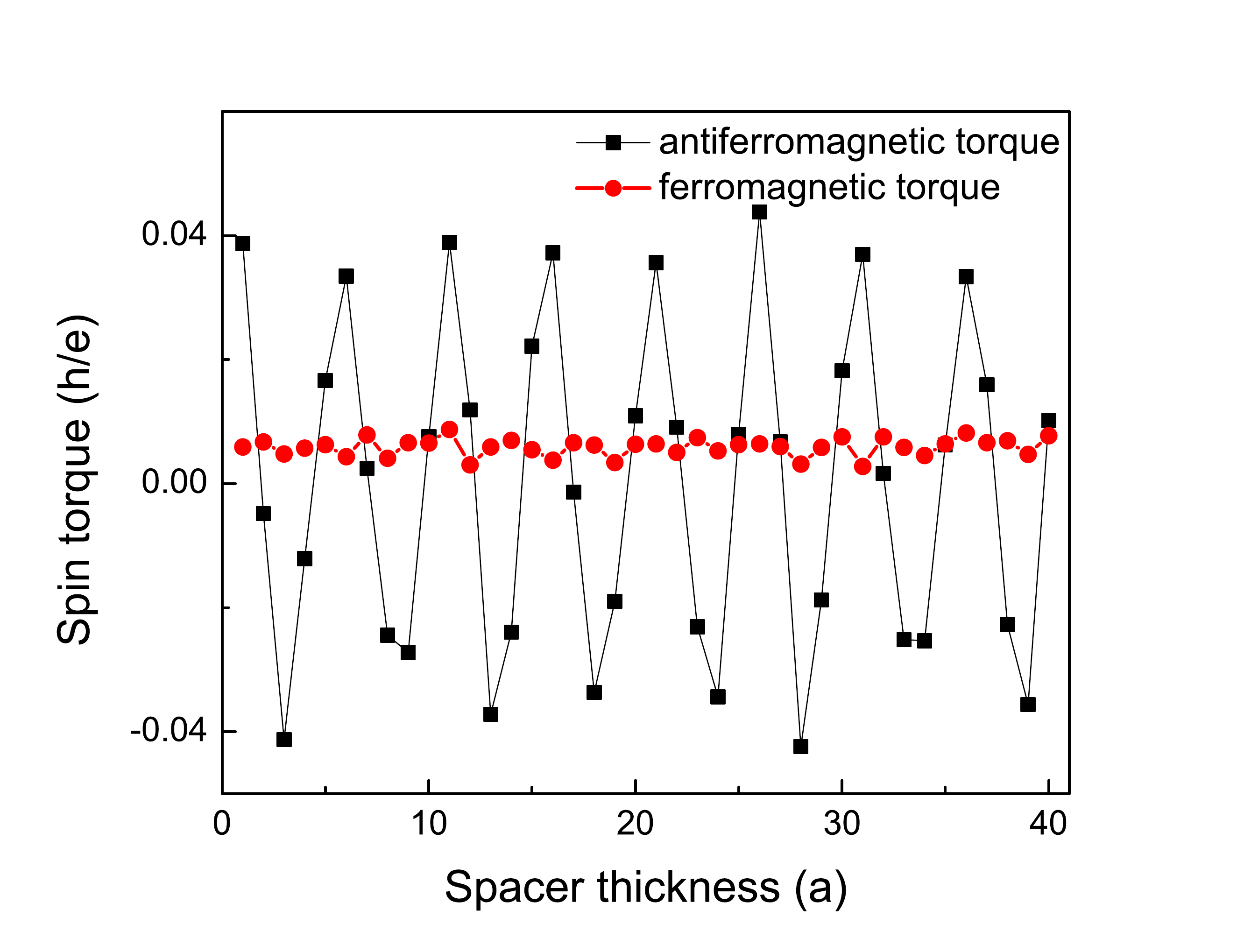}    	
  \caption{(Color online) The efficiency of the spin torque calculated for different spacer thicknesses in the case of a symmetric G-type antiferromagnetic spin-valve ($\Box$) and in the case of a symmetric ferromagnetic spin-valve ($\bigcirc$). The relative angle between the order parameters (magnetizations) of the two antiferromagnetic (ferromagnetic) layers equals to $\pi/2$. The calculations are performed in the absence of disorder, and the thicknesses vary from one to 40 layers. The parameters are the same as in Fig. \ref{fig:fig2}.\label{fig:fig6}}
  \end{center}
\end{figure}
Finally an important feature is the dependence of the spin torque magnitude as a function of the thickness of the metallic spacer, shown in Fig. \ref{fig:fig6}. In ballistic ferromagnetic spin-valves, it is well known that the spin torque and giant magnetoresistance oscillate with the thickness of the spacer due to spin-dependent reflections \cite{edwards}. However, the overall magnitude of the spin torque is only weakly affected (red dots in Fig. \ref{fig:fig6}). The resulting spin torque is therefore quite robust against spacer thickness variations as well as growth modulations. However, in the case of an antiferromagnetic spin-valve, the torque changes sign when varying the spacer layer thickness due to the quantum interferences between spin-dependent wave functions in the spacer. The oscillatory behavior reported in Fig. \ref{fig:fig6} indicates the extreme sensitivity of the antiferromagnetic spin torque to material parameters variation, in sharp contrast with the robustness of the ferromagnetic spin torque. It is therefore expected that small modifications of the system such as interfacial roughness or defects dramatically alter the spin torque in antiferromagnetic spin-valves.
\subsection{Spin torques in the presence of disorder}

\begin{figure}
\centering
\begin{tabular}{cc}{\bf (a)} &{\bf (b)}\\
\includegraphics[width=0.45\columnwidth]{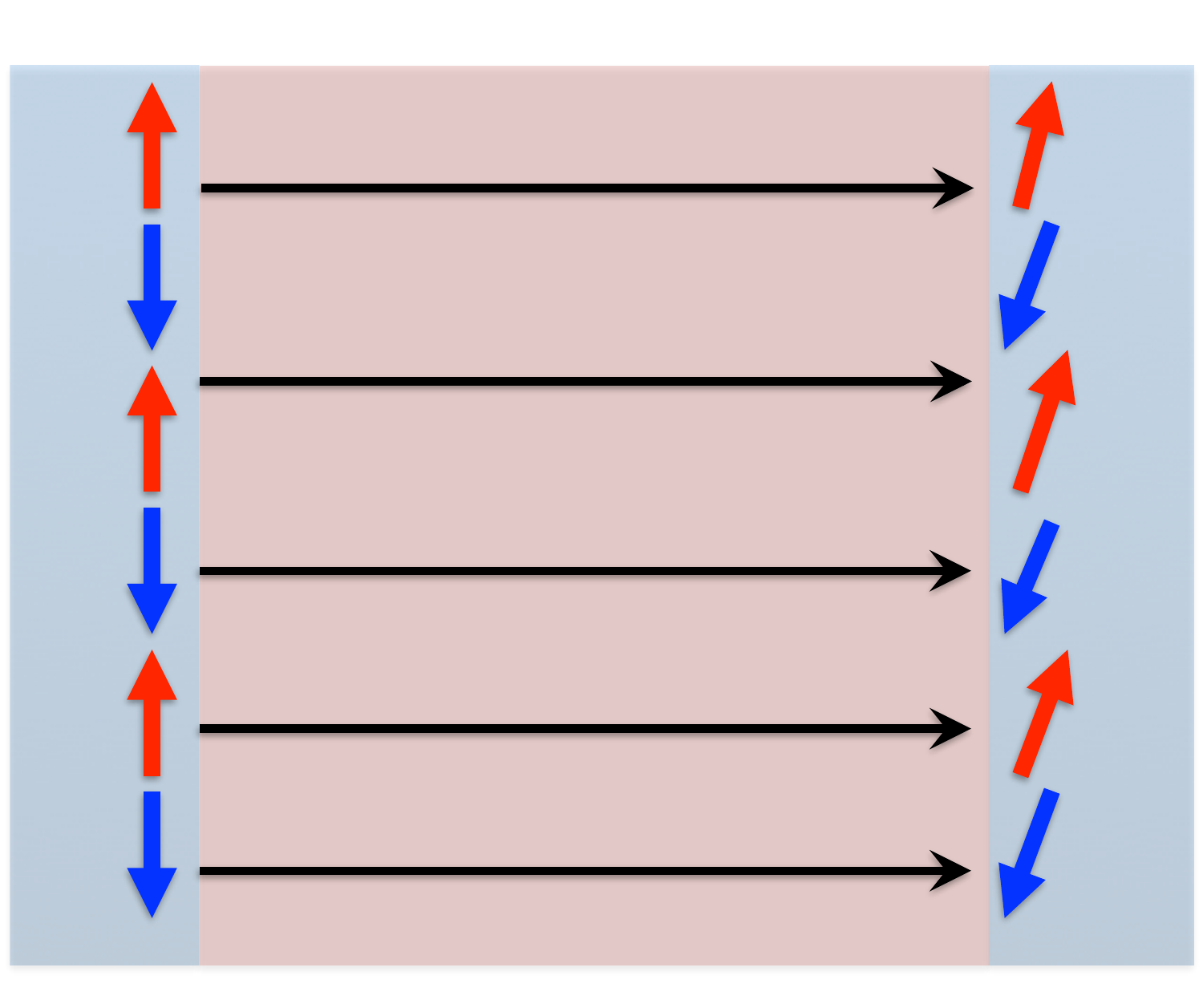} &
\includegraphics[width=0.45\columnwidth]{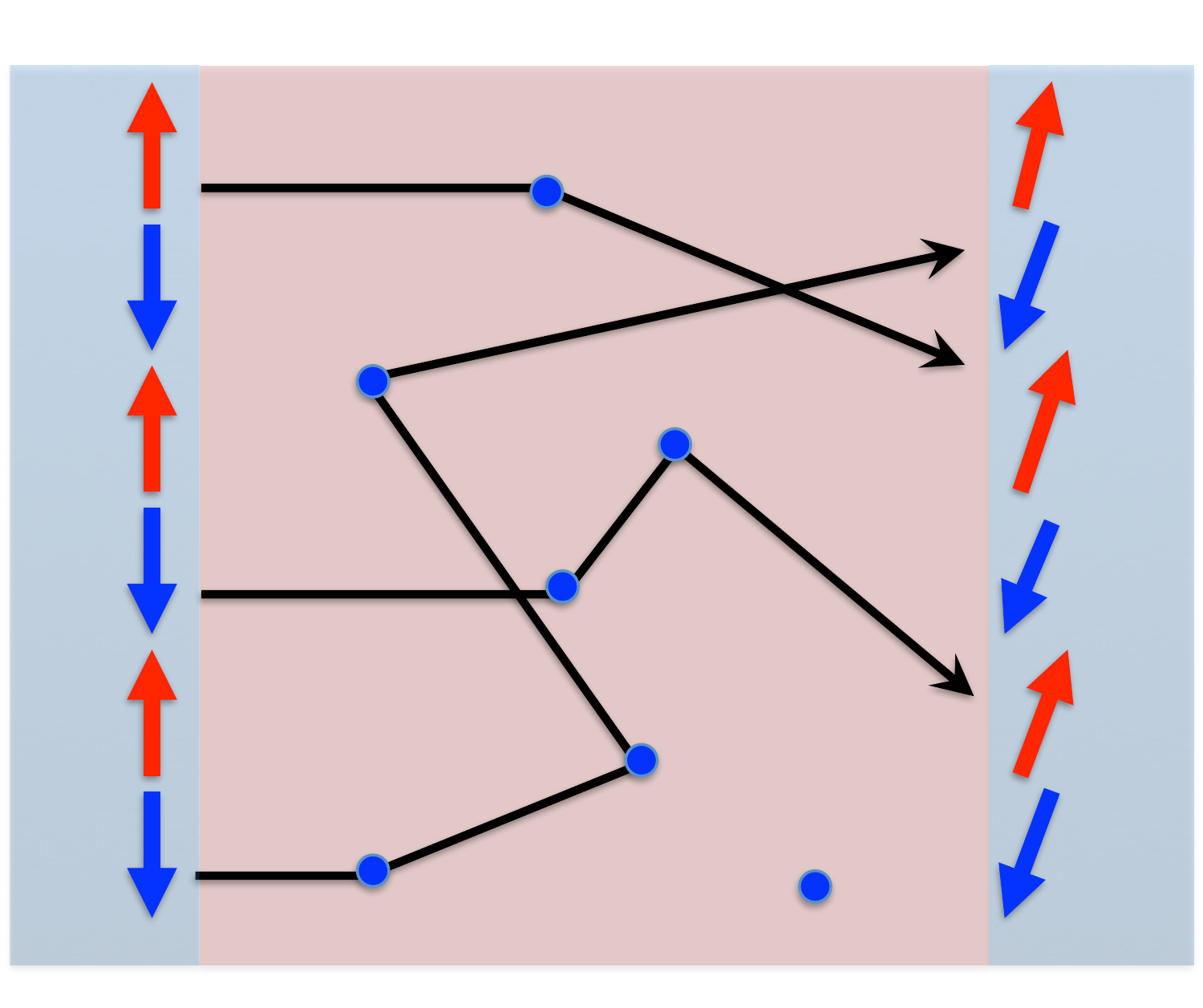}
\end{tabular}
\caption{(Color online) Schematics of the ballistic transport occurring in an antiferromagnetic spin-valve in absence (a) and presence (b) of disorder. In the former case, the direction of the momentum is not modified in the spacer. In the latter case, disorder scattering alters the linear momentum direction so that the staggered spin density built in the left antiferromagnet is redistributed over the right antiferromagnet, resulting in a dramatic reduction of the spin torque magnitude.\label{fig:figscat}}
\end{figure}
The results presented in the previous section assumed a ballistic transport throughout {\em disorder-free} spin-valves, as illustrated in Fig. \ref{fig:figscat}(a). Although this approach is expected to provide the correct torque angular dependencies, bulk momentum scattering due to disorder and impurities is an important ingredient of spin transport in metallic systems. More specifically, the large torque obtained in antiferromagnetic spin-valves is associated with the transmission of the spatial dependent staggered spin texture from one layer to another \cite{Nunez}. The coherence of this staggered spin texture is a crucial factor to obtain large antiferromagnetic torques that extend over the volume of the free layer, as already mentioned in Refs. \onlinecite{Nunez,duine2007}. In the presence of disorder, itinerant electrons are scattered in the spacer, resulting in a redistribution of the spin density impinging on the right layer, as illustrated in Fig. \ref{fig:figscat}(b). Therefore, one expects that momentum scattering is detrimental to the spin torque in 
antiferromagnetic spin-valves. In this section, we 
artificially introduce momentum scattering in the bulk of the layers to test the robustness of the previous results obtained in the clean limit.\par

\begin{figure}
\centering
\includegraphics[width=0.75\columnwidth]{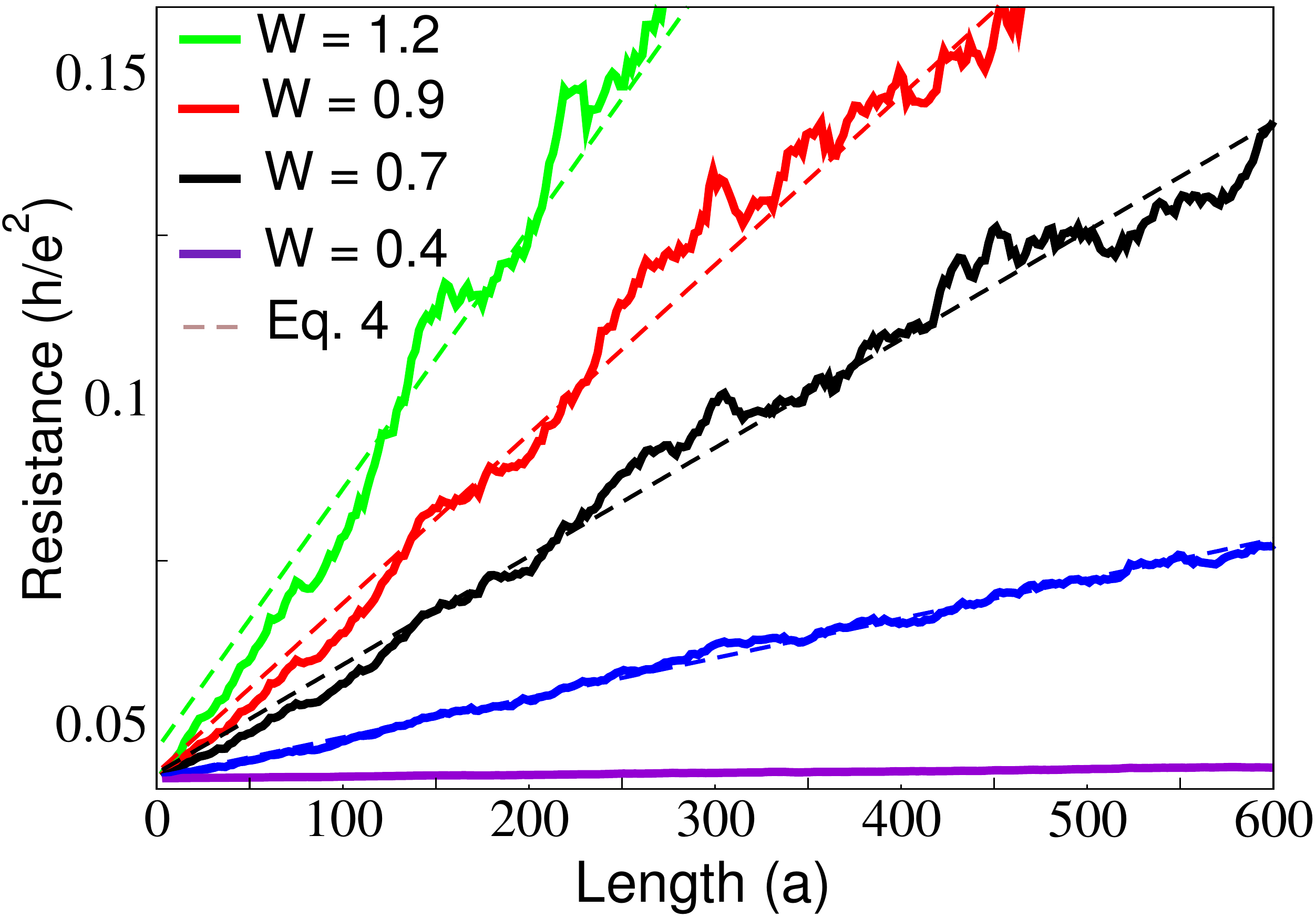}    	
\caption{(Color online) Resistance of a metallic layer as a function of the width of the layer for different disorder strengths W. The approximate linear relationship between the resistance and the width allows for extracting the effective mean free path of the metallic layer.\label{fig:fig7}}
\end{figure}

A model of bulk disorder is obtained by randomizing the on-site energy of the atoms $\epsilon_i$ over an energy range [-W,+W]. To map this approach to the equivalent mean free path $\lambda$, the resistance of each layer is calculated as a function of the layer thickness for different disorder strengths W, as shown in Fig. \ref{fig:fig7}. From the curves, we extract the mean free path for each disorder following the semi-classical formula of the conductance:
\begin{equation}
G=G_0/(1+\frac{L}{\lambda})
\end{equation}
$G_0=(e^2/h)N$ where $N$ stands for the number of transport channels in the sample. Depending on the disorder strength W, the mean free path varies for our calculation from 50 to 2500 atomic sites (equivalent to a range of 15 to 750 nm for a lattice parameter of $a_0=$0.3 nm). We verified that localization effects are negligible here. In order to calculate the torques in the disordered regime, we performed the calculation over 5000 disorder configurations and average over these configurations.\par

\begin{figure}
\centering
\includegraphics[width=1.\columnwidth]{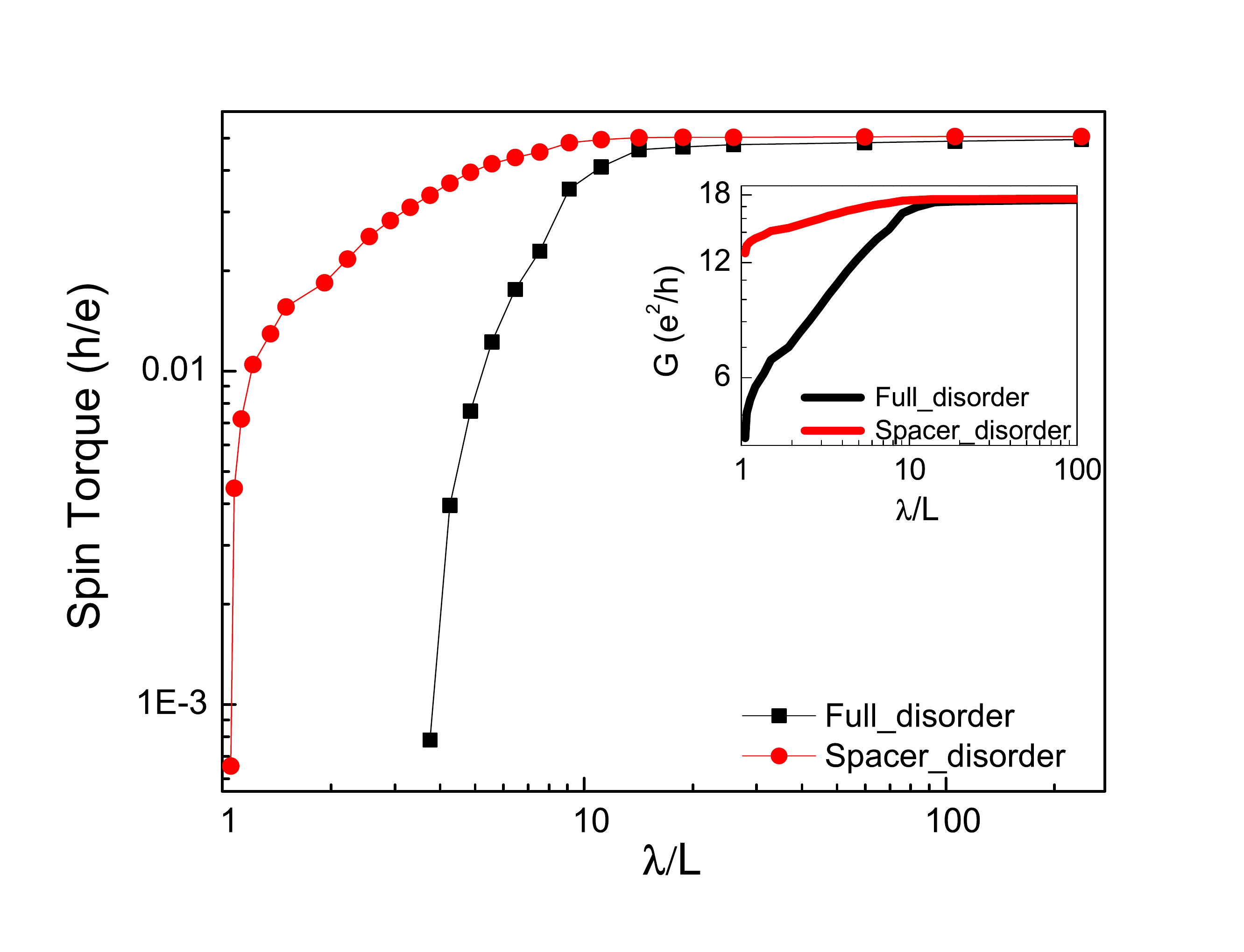}
\caption{(Color online) Spin torques as function of the reduced mean free path $\lambda/L$ calculated in fully disordered ($\Box$) and spacer-only disordered ($\bigcirc$) systems. The relative angle between the order parameters of the two antiferromagnetic layers equals to $\pi/2$. The inset shows the corresponding conductances.\label{fig:fig8} The calculation is averaged over 2500 configurations and the parameters are the same as in Fig. \ref{fig:fig2}. The logarithmic scale is used for both axis.}
\end{figure}
We first calculate the in-plane coherent torque in a symmetric G-type spin-valve when introducing disorder in the spacer only and in the whole spin-valve. Figure \ref{fig:fig8} shows the two torques as a function of the mean free path extracted from Fig. \ref{fig:fig7}. The inset displays the associated conductance, for reference. As expected, the presence of disorder throughout the whole spin-valve destroys the torque much faster than when disorder is only applied in the spacer. However in the remaining of this work, in order to reduce the computation time, we shall restrict ourselves to cases where disorder is present in the spacer layer only.\par

\begin{figure}
\centering
\includegraphics[width=1.\columnwidth]{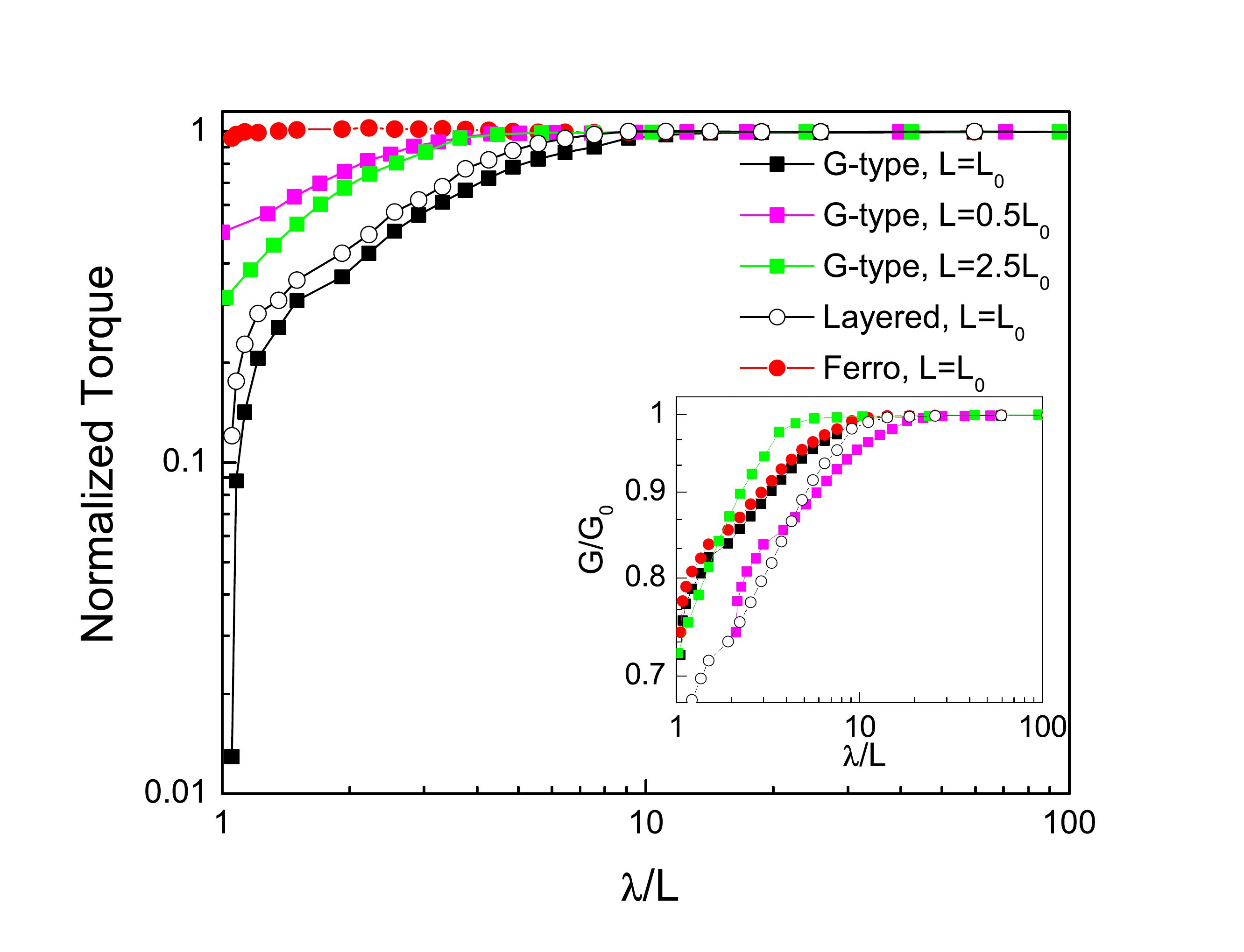}	
\caption{(Color online) Normalized torques as function of the reduced mean free path $\lambda/L$ calculated respectively in Ferromagnet (red), G-type Antiferromagnet with different thicknesses [black ($L_0$), magenta ($0.5L_0$) and green ($2.5L_0$)] and the L-type type (open circles) for a relative angle of $\pi/2$. The inset shows the corresponding conductances. The parameters are the same as in Fig. \ref{fig:fig2} and we use the logarithmic scale for both axis.\label{fig:fig9}}
\end{figure}

Figure \ref{fig:fig9} shows the spin torque in a symmetric G-type spin-valve with different thicknesses $L=\alpha L_0$ (square symbols)  as function of the mean free path (here $L_0=20$ sites). The torques are normalized to their values in the clean limit for a better comparison. The inset displays the conductance as a function of the normalized mean free path $\lambda/L$. We notice that neither the torques nor the conductances collapse within one single curve, as one could expect from the $1/(1+L/\lambda)$ scaling law. We attribute the absence of the scaling to the importance of the spin-dependent interferences uncovered in Fig. \ref{fig:fig6}. As a reference, we also show the spin torque in a ferromagnetic spin-valve with equivalent tight-binding parameters. This torque is very robust against disorder, whereas the torque in antiferromagnet, although formally larger (see Fig. \ref{fig:fig5}) in the clean limit, is dramatically reduced 
when turning on momentum scattering.\par

Finally, the torque in the L-type antiferromagnetic spin-valve is shown and displays similar features, almost overlapping with its G-type counterpart (open symbols). The extension of the disorder over the whole spin-valve (not shown) does not qualitatively change the results and the angular dependence of the torque is preserved in spite of the disorder.  The direct consequence of this observation is that it seems very difficult to maintain antiferromagnetic torques in conventional metallic systems. A possibility to overcome this obstacle would be to use ultra clean metallic systems (typically epitaxially grown stacks) or to even replace the metallic spacer by a tunneling barrier, as discussed in Section \ref{sec:disc}.

\subsection{Elastic versus inelastic scattering}
The present study shows without ambiguity that {\em elastic scattering} is detrimental to spin torque in antiferromagnets. It is instructive to compare the present work with  Duine et al. \cite{duine2007} which addresses the nature of spin torque in a 1-dimensional antiferromagnetic spin-valve using the Landauer-Buttiker formalism, in the same line as Ref. \onlinecite{Nunez}. Using a so-called voltage probe model, Ref. \onlinecite{duine2007} tested the impact of {\em inelastic scattering} (i.e. phase decoherence and spin relaxation) on the spin transport, magnetoresistance and spin torque. In this approach, the inelastic scattering (due to phonons or magnons) induces spin relaxation, while the momentum is, by definition of the Landauer-Buttiker formalism, conserved. In contrast, our approach applies to a two-dimensional system and captures elastic scattering only while keeping the transport phase coherent. Spin relaxation can be introduced in our code by implementing a random spin-dependent on-site energy or by implementing spin-orbit coupling in the hopping parameter, which is beyond the scope of the present study. Therefore, the present study and Ref. \onlinecite{duine2007} address two different mechanisms that should be equally present in realistic systems.

Interestingly, Ref. \onlinecite{duine2007} showed that the spin relaxation has a different impact on spin torques in ferromagnetic and antiferromagnetic spin-valves. In the former case, the torque efficiency is decreased whereas in the latter case, the torque efficiency increases (see Figs. 4 and 7 in Ref. \onlinecite{duine2007}). Notice that the magnitude of the change is no more than a factor 2 and the increase in torque efficiency is attributed to a decrease in the conductance (the staggered spin density is progressively destroyed which results in a conductivity reduction). These results contrast with the present work. Elastic scattering modeled through disorder has a much more dramatic impact on spin torque in antiferromagnet spin-valves than in ferromagnetic spin-valves (see Fig. \ref{fig:fig9}). The former is reduced by orders of magnitude whereas the latter is robust against disorder. Therefore, it seems that more than phase-coherent transport, momentum conservation is the key to achieve sizable spin torque in antiferromagnetic spin-valves.

\subsection{Materials considerations\label{sec:disc}}

We conclude this article by a few comments on the realization of spin transfer torque in realistic antiferromagnetic structures. Up till now, antiferromagnets based on heavy metals such as PtMn and IrMn have been widely used for their exchange bias properties. These materials are usually deposited using sputtering techniques, resulting in small grains \cite{stefanita} (from 5 to 15nm wide) which defines the electron mean free path and affects the interfacial properties. Indeed, to the best of our knowledge, short spin relaxation lengths of about 2 nm have been deduced from giant magnetoresistance \cite{bass,delille} and spin pumping studies \cite{merodio}. As a reference, our calculations reported in Fig. \ref{fig:fig8} indicate that the torque is reduced by one order of magnitude for a mean free path of $\lambda\approx 20$ atomic units ($\equiv$6 nm) when the disorder is in the spacer only and of $\lambda\approx 80$ atomic units ($\equiv$24 nm) when the disorder is everywhere. This clearly points out the major challenge paused by the controlled growth of clean antiferromagnetic spin-valves. In addition, the presence of interfacial disorder results in a broadening of the blocking temperature distribution \cite{baltz} that is detrimental for the (current-driven) coherent control of the antiferromagnetic order. Notice that lowering the temperature should result in a reduction of inelastic scattering, but not in the reduction of the defect- and impurity-induced mean free path. Hence, reducing the disorder seems to be a mandatory pathway towards antiferromagnetic spintronics. The recent realization of spintronics devices based on antiferromagnets deposited by sputtering techniques \cite{Wunderlich} or by epitaxial growth \cite{marti1,marti2}, as well as the fabrication and characterization of novel antiferromagnets such as Au$_2$Mn \cite{wu, barthem} and CuMnAs \cite{maca} offer interesting perspectives for spin transport in clean antiferromagnets with long spin diffusion length.

\section{Conclusion and perspectives\label{sec:4}}
The nature of spin transfer torque in antiferromagnetic structures has been investigated using the non-equilibrium Green's functions formalism. In particular, we considered symmetric spin-valves composed of G-type and L-type antiferromagnets, as originally proposed by N\'unez et al.\cite{Nunez}, as well as asymmetric spin-valves composed of a combination of G-type and L-type antiferromagnets. Our results show that (i) the torque in symmetric antiferromagnetic spin-valves consists of an in-plane coherent torque and a perpendicular exchange torque. (ii) The torque in asymmetric spin-valves consists of an perpendicular coherent torque and an in-plane exchange torque for the torque exerted on the G-type antiferromagnet, whereas it shows both in-plane and perpendicular coherent torque when exerted on the L-type antiferromagnet. (iii) Within the parameters used in the present work, the torques in antiferromagnetic spin-valves are found to be at least one order of magnitude larger than the torque calculated in 
ferromagnetic spin-valves. Finally, (iv) we demonstrated that momentum scattering is dramatically detrimental to spin torque in antiferromagnets. This indicates that the transport coherence is seminal to achieve sizable torques. In order to get rid of this constraint, one needs either a system where momentum conservation is ensured such as tunneling junctions. New structures involving both spin-orbit coupling and tunneling barriers such as the one studied by Park et al. \cite{Wunderlich} and Marti et al.\cite{marti1,marti2} open promising directions towards this goal.\par
The authors acknowledge fruitful discussions with V. Baltz, R.A. Duine, A. Brataas, T. Jungwirth and J. Sinova. X.W. was supported by the ERC Consolidator Grant MesoQMC.\par

\end{document}